\documentclass[aps,prl,reprint,amsmath,amssymb,superscriptaddress,showpacs,nobibnotes]{revtex4-1}

\usepackage{graphicx}
\usepackage{bm}     
\usepackage{hyperref} 
\usepackage{dsfont}    
\usepackage{color}
\usepackage[normalem]{ulem}
\usepackage{cleveref}

\newcommand{\ket}[1]{|#1\rangle}

\newcommand{\abs}[1]{\left|{#1}\right|}

\newcommand{\eq}[1]{Eq.~(\ref{#1})}
\newcommand{\fig}[1]{Fig.~\ref{#1}}

\def\beq{\begin{eqnarray}}
\def\eeq{\end{eqnarray}}

\def\A{\mathcal{A}}

\renewcommand{\section}[1]{{\textit{#1}.---}}

\begin{document}

\title{Reversed interplay of quantum interference and which-way information in multi-photon entangled states}


\author{Young-Sik Ra}
\altaffiliation{Present address: Laboratoire Kastler Brossel, Universit\'e Pierre et Marie Curie, 75252 Paris, France}
\email{youngsikra@gmail.com}
\affiliation{Department of Physics, Pohang University of Science and Technology (POSTECH), Pohang, 790-784, Korea}

\author{Malte C. Tichy}
\affiliation{Department of Physics and Astronomy, University of Aarhus, 8000 Aarhus C, Denmark}

\author{Hyang-Tag Lim}
\altaffiliation{Present address: Institute of Quantum Electronics, ETH Zurich, CH-8093 Zurich, Switzerland}

\affiliation{Department of Physics, Pohang University of Science and Technology (POSTECH), Pohang, 790-784, Korea}

\author{\mbox{Clemens Gneiting}}
\altaffiliation{Present address: RIKEN, Wako-shi, Saitama 351-0198, Japan}
\affiliation{Physikalisches Institut, Albert-Ludwigs-Universit\"at, Hermann-Herder-Str.~3, D-79104 Freiburg, Germany}

\author{Klaus M\o lmer}
\affiliation{Department of Physics and Astronomy, University of Aarhus, 8000 Aarhus C, Denmark}

\author{Andreas Buchleitner}
\affiliation{Physikalisches Institut, Albert-Ludwigs-Universit\"at, Hermann-Herder-Str.~3, D-79104 Freiburg, Germany}

\author{Yoon-Ho Kim}
\email{yoonho72@gmail.com}
\affiliation{Department of Physics, Pohang University of Science and Technology (POSTECH), Pohang, 790-784, Korea}

\date{\today}

\begin{abstract} 

We report experimental studies of the multi-photon quantum interference of a two-mode three-photon entangled Fock state $\ket{2,1}+\ket{1,2}$ impinging on a two-port balanced beam splitter.  When the distinguishability between the two input paths is increased, we observe a reduction followed by a resurgence of the quantum interference signal. We ascribe this unusual behavior to the competition among contributions from distinct numbers of interfering photons. Our theoretical analysis shows that this phenomenon will occur for any entangled Fock-state input of the form $\ket{N,M}+\ket{M,N}$ where $M,N>0$. Our results are an indication that wave-particle duality may give rise to a wide range of, largely unexplored, phenomena in multi-particle interference.
\end{abstract}

%

\maketitle

Quantum interference is one of the most fundamental features of quantum mechanics, observed in a variety of quantum systems~\cite{Grangier:1986aa,Hong:1987aa,Gerlich:2011go,Bocquillon:2013dp,Lopes:2015cn}. A prototype example is the double slit experiment, where the repeated incidence of a single particle leaves wave-like interference fringes on a screen~\cite{Feynman:1965aa,Grangier:1986aa}. Perfect interference is only observed if no information is available about which path the particle has taken through the slits ~\cite{Pittman:1996p10615}, while partial path distinguishability gradually reduces the fringe contrast~\cite{Jaeger:1995eg,Englert:1996aa}. For example, a time delay applied to one of the two paths yields which-path information and ultimately causes the  interference fringes to vanish. Previous studies on interference of a single photon and two photons have thus shown that increasing the distinguishability simply reduces the interference fringe visibility~\cite{Rarity:1990p6077,Durr:1998bn,Schwindt:1999hx,Hackermuller:2004p6499,Jacques:2008aa}, as a quantitative consequence of wave-particle duality~\cite{Jaeger:1995eg,Englert:1996aa}.

In this work, we study multi-photon interference and observe that, in contrast to the single-photon case, interference signals may vanish and reappear under a gradually \emph{increased} path distinguishability. Our theoretical analysis reveals that the observed phenomena are not due to information erasure~\cite{Kwiat:1992aa,Kim:2000aa}, non Markovian processes~\cite{Xu:2010aa,FRANCO:2013cr}, or a periodic decoherence~\cite{LoFranco:2012p10749,Xu:2013ia}, but due to a passage between different numbers of interfering photons which exhibit distinct interference fringes. We find that with the exception of $N00N$ states, multi-photon states in general exhibit a nontrivial relationship between interference fringes and which-path distinguishability.


\begin{figure}[b]
\centerline{\includegraphics[width=3.3in]{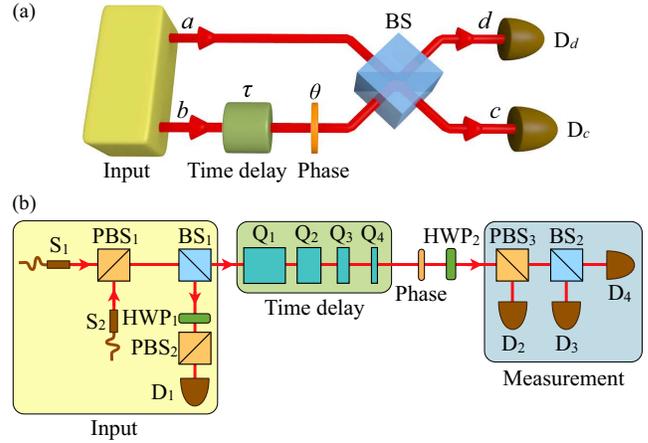}}
\caption{(a) Schematic of interferometer. BS is a non-polarizing balanced beam splitter. D$_c$ and D$_d$ detect $m$ and $n$ photons, respectively, defining ($m$,$n$) detection. (b) Experimental setup. The path modes $a$ and $b$ are realized by horizontal and vertical polarization modes. Half wave plates (HWP$_1$, HWP$_2$) are at 22.5$^\circ$. Quartz plates Q$_1$, Q$_2$, Q$_3$, and Q$_4$ have different thicknesses 7$l$, 4$l$, 2$l$, and $l$ (= 1.7 mm), respectively.
The phase on the vertical polarization is controlled by rotating an HWP between two quarter wave plates at 45$^\circ$ (not shown).
D$_1$$\sim$D$_4$ are single-photon detectors. HWP$_2$ plays the role of BS in (a).}\label{fig:setup}
\end{figure}

\begin{figure*}[t]
\centerline{\includegraphics[width=0.75\textwidth]{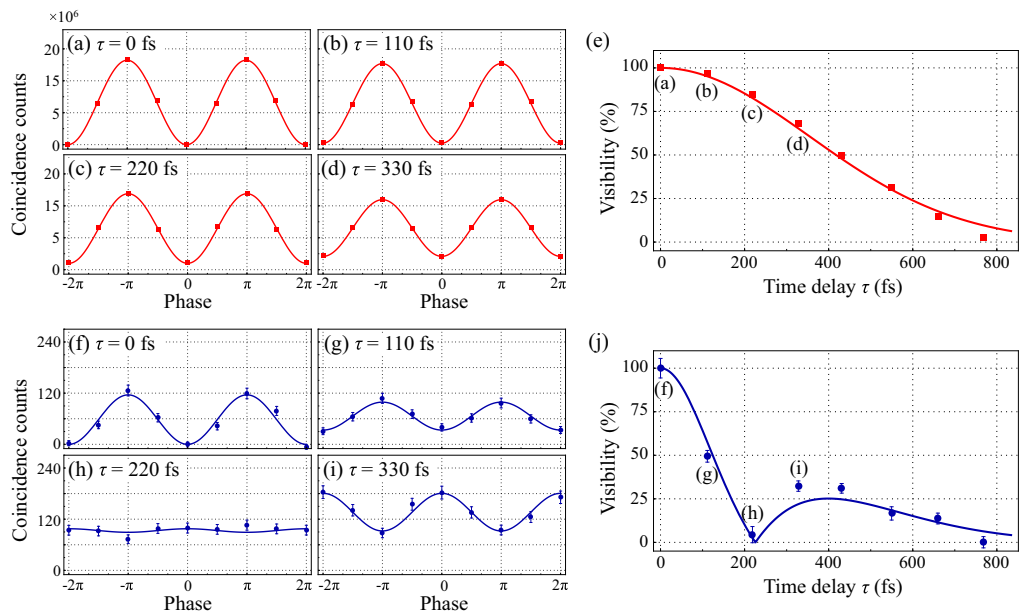}}
\caption{Interference at various time delays. Single-photon interference fringes by $\ket{1,0}_{H,V}+\ket{0,1}_{H,V}$ are shown in (a-d), and the visibility is summarized in (e). Three-photon interference fringes by $\ket{2,1}_{H,V}+\ket{1,2}_{H,V}$ are shown in (f-i), and the visibility is summarized in (j). Red squares and blue dots are experimental data. For the single-photon interference, coincidence counts on D$_1$ and D$_2$ are recorded, and for the three-photon interference, coincidence counts on D$_1\sim$D$_4$ are recorded. Each of data points in (a-d,f-i) is accumulated for 6600 s, and noise counts estimated by an additional photon-pair generation are subtracted~\cite{Liu:2008p10810}. The error bars represent one standard deviation. The curves in (a-d,f-i) are sinusoidal fits of the experimental data. The red and blue curves in (e) and (j) are fits of the experimental data to the theoretical model in \eq{eq:visibilitysingle} and \eq{eq:visibilitythree}.
}\label{fig:data}
\end{figure*}

\section{Experimental results}
Figure \ref{fig:setup}(a) shows a schematic of our experimental setup to observe multi-photon interference.
This setup, illuminated with the single-photon state
\beq
\ket{\psi_{1:0}}=\frac{1}{\sqrt{2}}\left( \ket{1,0}_{a,b}+\ket{0,1}_{a,b} \right),
 \label{eq:single-photon} \eeq
is equivalent to the aforementioned double-slit experiment~\cite{Feynman:1965aa,Grangier:1986aa}, and single-photon detection at D$_d$, denoted as (0,1) detection, shows interference fringes as the interferometer phase $\theta$ is varied.
When illuminated by a three-photon entangled state
\beq
\ket{\psi_{2:1}}=\frac{1}{\sqrt{2}}\left( \ket{2,1}_{a,b}+\ket{1,2}_{a,b} \right),
 \label{eq:three-photon} \eeq
 we observe interference fringes in the coincidence detection of two photons at D$_c$ and a single photon at D$_d$, denoted as (2,1) detection, while scanning the phase $\theta$.  In both the single- and three-photon experiments, the distinguishability between paths $a$ and $b$ can be tuned by introducing a time delay $\tau$ in path $b$: as $\tau$ increases, the path distinguishability becomes larger because the arrival time of a photon at the beam splitter (BS) provides which-path information~\cite{Jaeger:1995eg,Englert:1996aa,Schwindt:1999hx,Ra:2013ab}.

In the experiment, we implement the interferometer by exploiting the horizontal and vertical polarization modes of a photon, as sketched in \fig{fig:setup}(b). To prepare the single-photon state in~\eq{eq:single-photon} and the three-photon state in~\eq{eq:three-photon}, we use photon pairs generated by spontaneous parametric down conversion via type-I non-collinear phase matching (not shown in \fig{fig:setup}(b)): a femtosecond pulse laser (duration: 150 fs, repetition rate: 95 MHz, central wavelength: 390 nm, average power: 190 mW) pumps a 2 mm thick $\beta$-BaB$_2$O$_4$ crystal, where each of the generated photons is spectrally filtered by a narrow band-pass filter (full width at half maximum of 3 nm) and spatially filtered by coupling into a single-mode fiber (S$_1$ or S$_2$ in \fig{fig:setup}(b)), which ensures indistinguishability among the generated photons. Photons from S$_1$ and S$_2$ are horizontally ($H$) and vertically ($V$) polarized, respectively, and arrive simultaneously at a polarizing beam splitter (PBS$_1$). The quantum state of the photons after PBS$_1$ is $(1-\abs{\eta}^2)^{1/2} \ \sum_{n=0}^{\infty} \eta^n\ket{n,n}_{H,V},$ where we exploit the single-pair-term $\ket{1,1}_{H,V}$ and the two-pair-term $\ket{2,2}_{H,V}$ by detecting coincidence counts at D$_1$ and D$_2$ and D$_1 \sim$ D$_4$ (Perkin-Elmer SPCM-AQRH-13), respectively. To avoid contributions from higher-order photon-pair generation, we use  low pump power (190 mW), which gives $\abs{\eta}^2=0.02$. From the single-pair-term (the two-pair-term), the single-photon state in~\eq{eq:single-photon} (the three-photon state in~\eq{eq:three-photon}) is prepared by detecting a single photon at D$_1$~\cite{Ra:2016jh}. Time delays are implemented on the vertical polarization mode by using four different-thickness quartz plates (Q$_1\sim$Q$_4$), yielding time delays of $0$, $\tau_0$ (=110 fs), $\dots, $ $7 \tau_0$. For each time delay, we record single-photon interference fringes by detecting coincidence counts at D$_1$ and D$_2$ as well as three-photon interference fringes by detecting coincidence counts at D$_1 \sim $D$_4$ while scanning the interferometer phase $\theta$.

For single-photon interference, we observe a gradual reduction of the interference fringes in \fig{fig:data}(a-d) with increased time delays, summarized in \fig{fig:data}(e). The straightforward and monotonic relation between the time delay and the visibility agrees well with the wave-particle duality relation~\cite{Jaeger:1995eg,Englert:1996aa}. In three-photon interference, however, we encounter in \fig{fig:data}(f-i) a qualitatively different behavior of the interference fringes, and, in particular, the three-photon interference vanishes at $\tau= 220$ fs (\fig{fig:data}(h)) and reappears (with a $\pi$-phase shift) at a further increased time delay (\fig{fig:data}(i)). The behavior of the fringe visibility is summarized in \fig{fig:data}(j).


\section{Theoretical analysis}
To explain why the three-photon interference exhibits the observed nontrivial behavior, we apply a multimode analysis for the multi-photon state. The  creation operator for a photon occupying a Gaussian wave packet centered at time $t$ can be described as
\beq
\A^\dagger=\frac{1}{\sqrt \pi\Delta \omega} \int \mbox{d} \omega \exp\left(- \frac{(\omega-\omega_0)^2}{2 \Delta\omega^2} + i \omega t \right)  \A_{\omega}^\dagger, \label{eq:creationoper}
\eeq
where $\A_\omega^\dagger$ is the creation operator of a photon with definite frequency $\omega$, and $\omega_0$ ($=2.41 \times 10^{15}~\text{s}^{-1}$) and $\Delta\omega$($=3.99 \times 10^{12}~\text{s}^{-1}$) are the central frequency and the bandwidth, respectively. The operator $\A^\dagger (\tau)$, creating a single photon in the wave packet with a time delay $\tau$, has a similar expression, and it can be expanded on the creation operator without delay $\A_0^\dagger$ and an orthogonal component, readily found by Gram-Schmidt orthonormalization~\cite{Tichy:2011p10582,Ra:2013ab,Ra:2013aa}:
\beq
\A^\dagger (\tau) = \alpha \A_0^\dagger + \beta \A_1^\dagger, \label{eq:delayedcreationoper}
\eeq
where $\A_1^\dagger$ is the creation operator of the orthonormalized mode, and
\beq
\alpha &=& e^{i \theta} \exp\left({-\Delta\omega^2 \tau^2 /4}\right), \nonumber \\
\beta &=& e^{i \theta} \sqrt{1-\exp\left({-\Delta\omega^2 \tau^2 /2}\right) } \label{eq:alpha_beta},
\eeq
with $\theta=\omega_0 \tau$ and $\abs{\alpha}^2+\abs{\beta}^2=1$. The delay $\tau$ reduces $\abs{\alpha}^2$ (consequently, $\abs{\beta}^2$ increases), and it thus induces a transition of $\A^\dagger (\tau)$ from $\A_0^\dagger$ to $\A_1^\dagger$.

\begin{figure}[t]
\centerline{\includegraphics[width=0.48\textwidth]{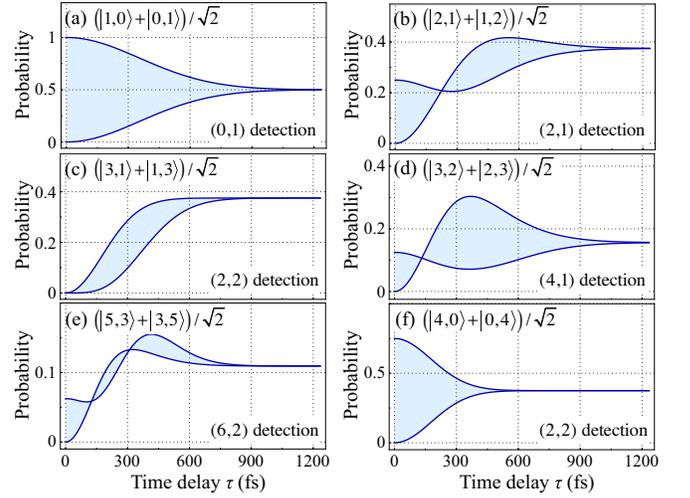}}
\caption{Detection probabilities of input states versus a time delay.
Shaded areas represent interference fringes with a period of $\omega_0$ for (a,b,d), $2\omega_0$ for (c) and (e), and $4\omega_0$ for (f).
}\label{fig:theory}
\end{figure}

Based on this decomposition, the single-photon state in \eq{eq:single-photon} exposed to a time delay $\tau$ in path $b$ of the interferometer, is described as
\beq
\ket{\psi_{1:0} (\tau)}&=&\left( \frac{1}{\sqrt{2}} \ket{1,0}_{a,b}+\frac{\alpha}{\sqrt{2}} \ket{0,1}_{a,b}\right)+\frac{\beta}{\sqrt{2}} \ket{0,\widetilde{1}}_{a,b}, \nonumber \\
 \label{eq:delayed_single-photon} \eeq
where a photon number without (with) tilde denotes photons created by $\A_0^\dagger$ ($\A_1^\dagger$). The first two terms interfere at the BS, but the last term does not interfere with the first two as it describes a photon occupying an orthogonal mode. The time delay then induces a gradual transition from single-photon interference ($\abs{\alpha}^2=1$, $\abs{\beta}^2=0$) to no interference ($\abs{\alpha}^2=0$, $\abs{\beta}^2=1$). The detection probability at D$_d$ is given by
\beq
P_{(0,1)} (\tau) = \frac{1}{2} \left( 1 - \abs{\alpha} \cos \theta \right), \label{eq:probsingle}
\eeq
which yields a visibility of
\beq
V_{(0,1)}(\tau) = \abs{\alpha}. \label{eq:visibilitysingle}
\eeq
The detection probability is plotted in \fig{fig:theory}(a), which shows a gradual degradation of interference fringes as expected from the wave-particle duality relation~\cite{Jaeger:1995eg,Englert:1996aa}.

\begin{figure}[t]
\centerline{\includegraphics[width=0.40\textwidth]{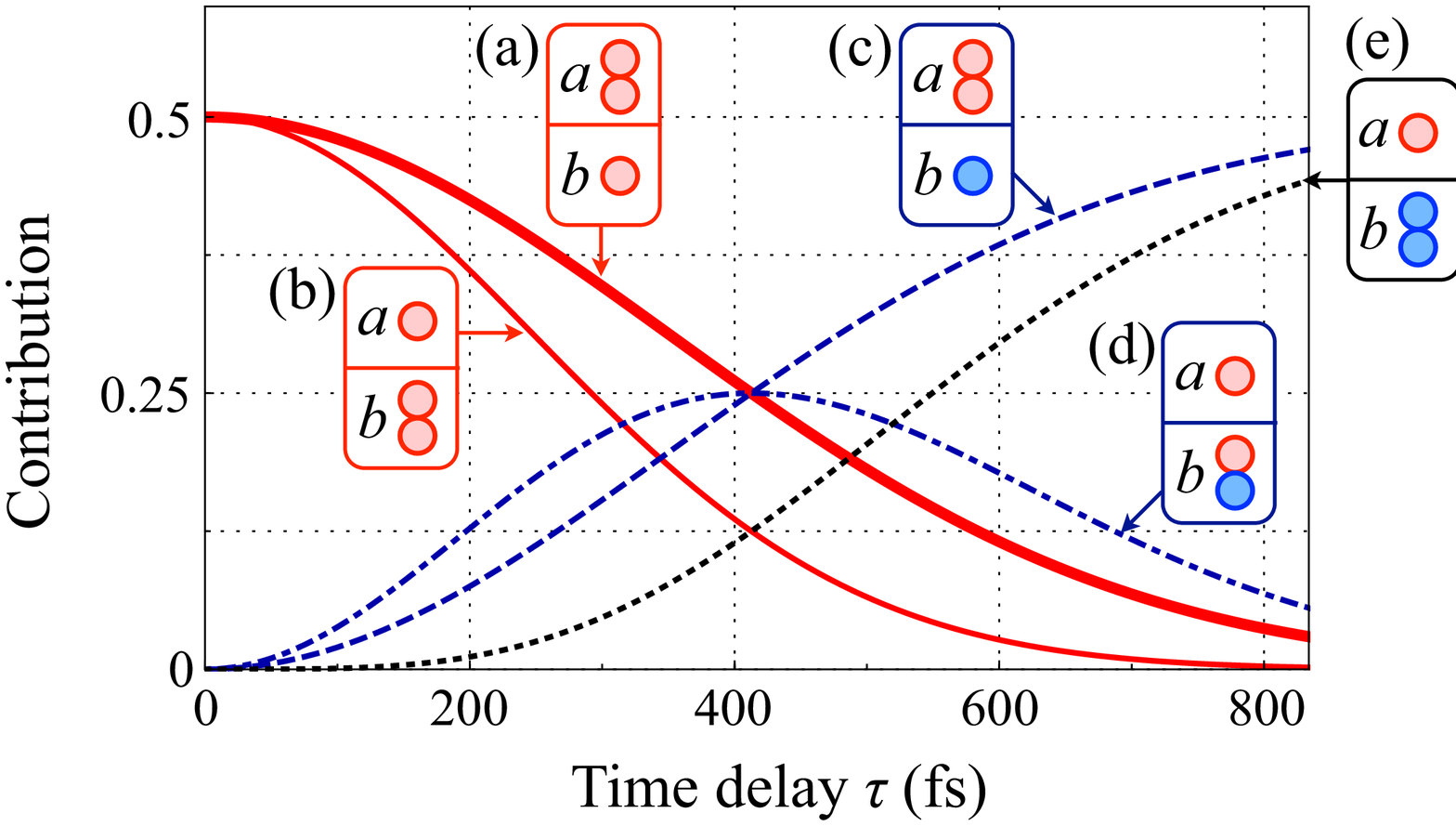}}
\caption{
Contribution of each term in \eq{eq:delayed_three-photon}. (a) $\ket{2,1}_{a,b}$ and (b) $\ket{1,2}_{a,b}$ can interfere at BS in \fig{fig:setup}(a). Similarly, (c) $\ket{2,\widetilde{1}}_{a,b}$ and (d) $\ket{1,1\widetilde{1}}_{a,b}$ can interfere. (e) $\ket{1,\widetilde{2}}_{a,b}$ is not involved in any interference. Red and blue disks represent photons created by $\A_0^\dagger$ and $\A_1^\dagger$, respectively.
}\label{fig:contribution}
\end{figure}

The three-photon state in \eq{eq:three-photon} is generated by the same creation operators and acquires a more complicated form when exposed to the time delay $\tau$,
\beq
\ket{\psi_{2:1}(\tau)}&=& \left( \frac{\alpha}{\sqrt{2}} \ket{2,1}_{a,b} + \frac{\alpha^2}{\sqrt{2}} \ket{1,2}_{a,b} \right) \nonumber \\
&+& \left( \frac{\beta}{\sqrt{2}} \ket{2,\widetilde{1}}_{a,b} + \alpha \beta \ket{1,1\widetilde{1}}_{a,b} \right) + \frac{\beta^2}{\sqrt{2}} \ket{1,\widetilde{2}}_{a,b}. \nonumber \\
 \label{eq:delayed_three-photon} \eeq
Here, qualitatively different interference types coexist: the first parenthesis represents interference of three indistinguishable photons; the second represents interference of only two indistinguishable photons (the third photon occupying the orthogonal mode in path $b$); the last term does not lead to any interference.
The three terms have different magnitude as the time delay increases: initially, $\ket{2,1}_{a,b}$ and $\ket{1,2}_{a,b}$ dominate, then  $\ket{1,1\widetilde{1}}_{a,b}$ and  $\ket{2,\widetilde{1}}_{a,b}$, and, finally  $\ket{1,\widetilde{2}}_{a,b}$ dominates the state, see \fig{fig:contribution}. As a result, the time delay displays a transition from three-photon interference over two-photon interference to no interference. The (2,1) detection probability, plotted in \fig{fig:theory}(b), is thus given by
\beq
P_{(2,1)} (\tau) &=& P^{\text{[three]}} (\tau) + P^{\text{[two]}} (\tau) + P^{\text{[no]}} (\tau),
\label{eq:probthree}
\eeq
where
\beq
P^{\text{[three]}} (\tau) &=& \abs{\alpha}^2 (\abs{\alpha}^2 - 2\abs{\alpha} \cos \theta+1)/16 \nonumber \\
P^{\text{[two]}} (\tau) &=& \abs{{\beta}}^2 (4 \abs{\alpha}^2 + 4 \abs{\alpha} \cos\theta + 3)/16 \nonumber \\
P^{\text{[no]}} (\tau) &=& 3\abs{\beta}^4/16,
\label{eq:eachdetprob}
\eeq
and the visibility becomes
\beq
V_{(2,1)}(\tau) = \frac{\abs{\alpha(2-3 \abs{\alpha}^2)}}{3-2\abs{\alpha}^2}. \label{eq:visibilitythree}
\eeq
We can now see that the vanishing and reappearance of interference fringes shown in \fig{fig:theory}(b) is due to the $\pi$ phase difference (see \fig{fig:data}(f-i)) between the three-photon and the two-photon interference fringes, $P^{\text{[three]}} (\tau)$ and $P^{\text{[two]}} (\tau)$. First the three-photon interference dominates, but for a critical delay, the three- and two-photon interference signals add to a constant, while for larger decay the two-photon interference dominates until, eventually, the detection probability is governed by the no-interference type $P^{\text{[no]}} (\tau)$.

\section{Discussion}
As we have seen, multi-photon interference exhibits a nontrivial dependence on the path distinguishability because of the different numbers of interfering photons contributing to the overall interference signal. Using our creation operators and their expansion on orthonormal modes, we can account for the time delay $\tau$ at mode $b$ for any linear combination of input Fock states on the interferometer. Let us for example consider the $(N+M)$-photon state $\ket{\psi_{N:M}}=\left( \ket{N,M}_{a,b}+\ket{M,N}_{a,b} \right) / \sqrt{2}$, where $N>M$. After the action of the time delay, this state can be written as
\beq
\ket{\psi_{N:M}(\tau)} &=& \frac{1}{\sqrt{2}} \sum_{k=0}^M \sqrt{M \choose k} \alpha^{M-k} \beta^k \ket{N,(M-k)\tilde{k}}_{a,b} \nonumber \\
&+& \frac{1}{\sqrt{2}} \sum_{l=0}^N \sqrt{N \choose l} \alpha^{N-l} \beta^l \ket{M,(N-l)\tilde{l}}_{a,b},~~~~~
 \label{eq:generalnm} \eeq
which leads to interference contributions  from states with total photon numbers in the ``non-tilded" modes created by $\A_0^\dagger$, ranging from $N$ to $N+M$. Figures \ref{fig:theory}(c-f) show different examples of interference fringes as a function of the time delay.
For a four-photon state $\left( \ket{3,1}_{a,b}+\ket{1,3}_{a,b} \right) / \sqrt{2}$, interference fringes by (2,2) detection is shown in \fig{fig:theory}(c).
At zero time delay, no interference fringe appears because neither $\ket{3,1}_{a,b}$ nor $\ket{1,3}_{a,b}$ can be detected by (2,2) detection~\cite{Ra:2013aa,Campos:1989vz}; however, when the time delay is introduced, interference fringes from three indistinguishable photons emerge. Another example shown in \fig{fig:theory}(d) is a five-photon state $\left( \ket{3,2}_{a,b}+\ket{2,3}_{a,b} \right) / \sqrt{2}$ measured by (4,1) detection. Similar to the three-photon state in \eq{eq:three-photon}, the five-photon state exhibits vanishing and reappearance of interference. An eight-photon state $\left( \ket{5,3}_{a,b}+\ket{3,5}_{a,b} \right) / \sqrt{2}$ exhibits a more complex behavior when measured by (6,2) detection: vanishing and reappearance of interference take place twice, shown in \fig{fig:theory}(e).
Remarkably, the states $\ket{\psi_{N:M}}$ with $M > 0$ in Figs. \ref{fig:theory}(c-e) show a larger tolerance to the path distinguishability than the $N00N$ state $\ket{\psi_{N:0}}$ in \fig{fig:theory}(f) which always shows a simple and rapid reduction of the fringe visibility, even though the former contains a larger or equal number of photons than the latter.


\section{Conclusion}
The observed vanishing and reappearance of multi-photon interference in the path distinguishability transition is due to contributions to the overall interference signal from different numbers of interfering photons, and the observation clearly demonstrates that straightforward application of the wave-particle duality relation~\cite{Jaeger:1995eg,Englert:1996aa} is not sufficient to account for multi-photon interference experiments. Our results, on the one hand, provide a new characteristic of multi-photon interference~\cite{Tichy:2014gm,Tillmann:2015bq,Tichy:2011p10582,Pan:2012kv,Ra:2013ab,Ra:2013aa,Jin:2016jf,Menssen:2016wi}, and, on the other hand, they may inspire investigation of a more foundational character, cf. the different view on wave-particle duality in first and second quantization~\cite{Jaeger:1995eg,Englert:1996aa,Huang:2013cg,Banaszek:2013iq}.

 From a practical  perspective, quantum technologies, such as precision measurements~\cite{Giovannetti:2011aa,Humphreys:2013gz,Tichy:2015jb,Ra:2015bb} and quantum simulations ~\cite{Tichy:2014gm,Tillmann:2015bq,Carolan:2015fb} are increasingly based on multi-photon interference and entanglement. It is, hence, pertinent to understand how these phenomena are affected by the nontrivial dependence on distinguishability.

This work was supported by the National Research Foundation of Korea (Grant Nos. 2016R1A2A1A05005202 and 2016R1A4A1008978), by the Danish
Council for Independent Research and by the Villum Foundation.





\clearpage

\end{document}